\documentclass{jpp}
\usepackage{graphicx}

\usepackage[utf8]{inputenc}
\usepackage[T1]{fontenc}
\usepackage{amsmath}
\usepackage{xcolor}

\makeatletter
\def\@underjournal{}
\makeatother

\title{Magnetic Field Line Chaos, Cantori, and Turnstiles in Toroidal Plasmas}

\author{Allen H. Boozer
  \corresp{\email{ahb17@columbia.edu}}
 }

\affiliation{
Columbia University, New York, NY 10027, USA}

\begin{document}

\maketitle

\begin{abstract}
\color{red} \color{black} The mathematical concepts of chaos, cartori, and turnstiles underlie a number of areas of tokamak and stellarator physics. Nevertheless, these concepts have seldom explicitly appeared in publications on fusion plasmas.  The absence of  physical intuition about these concepts is responsible for misunderstandings and slows developments in a number of areas: magnetic reconnection, the most important electromagnetic correction to what are called electrostatic microinstabilities, non-resonant divertors in stellarators, disruptions and damage from runaway electrons in tokamaks. Physicists become interested in new mathematical concepts when they give insights into and solutions to practical problems.  The importance of this review is not only in explaining chaos, cartori, and turnstiles as mathematical concepts but also in illustrating their significance through applications. \color{black}

\end{abstract}

\section{Introduction}

The importance of magnetic field line chaos, cantori, and turnstiles to understanding the behavior of toroidal plasmas has not yet led to familiarity with these concepts.  To increase that familiarity, these concepts will be defined and applications that illustrate their importance will be discussed.  

\begin{figure}
\centerline{ \includegraphics[width=3.2 in]{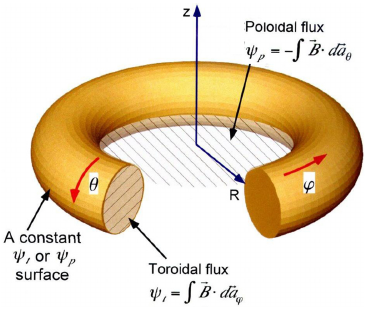}}
\caption{The poloidal flux $\psi_p$ is defined by the magnetic flux penetrating the hole in the center of the toroidal surface defined by a constant-$\psi_p$ surface.  The toroidal flux $\psi_t$ is defined by the magnetic flux enclosed by a toroidal surface defined by a constant-$\psi_t$ surface.  The poloidal $\theta$ and toroidal $\varphi$ angles can be chosen with arbitrariness.   This was Figure 1 in Boozer, Phys. Plasmas \textbf{26}, 122902 (2019).} 
\label{fig:flux}
\end{figure}

Magnetic field lines are defined by $d\vec{x}/d\ell = \vec{B}(\vec{x},t)/B$ with the time $t$ held fixed.  The parameter $\ell$ has the physical interpretation of the distance along the line.  In discussing their properties, several coordinate systems are needed.  What is meant by a coordinate system merits mention. 

Positions in space are fundamentally defined by Cartesian coordinates, $\vec{x}=x\hat{x}+y\hat{y}+z\hat{z}$.   Any three quantities $(\psi_t,\theta,\varphi)$ can be used as coordinates by defining the position function $\vec{x}(\psi_t,\theta,\varphi)$ as long as the Jacobian 
\begin{equation}
\mathcal{J}\equiv \left(\frac{\partial \vec{x}}{\partial\psi_t} \times \frac{\partial \vec{x}}{\partial\theta} \right) \cdot \frac{\partial \vec{x}}{\partial\varphi}
\end{equation}
is neither zero nor infinity in the spatial region in which they are to be used.  Defining the position function just means giving $x$, $y$, and $z$ as functions of $(\psi_t,\theta,\varphi)$.   The appendix to  \citep{Boozer:2004} explains how arbitrary coordinate systems, such as  $\vec{x}(\psi_t,\theta,\varphi)$, are used in calculations.

 One \color{black} reason coordinate systems are important to a discussion of the properties of magnetic field lines, is their trajectories are given by a Hamiltonian of the $H(p,q,t)$ form.  The Hamiltonian, its canonical momentum $p$ and canonical coordinate $q$, and its canonical time $t$ have a subtle relationship to physical quantities.  Section \ref{Sec:B-rep} shows the Hamiltonian is the poloidal flux $\psi_p(\psi_t,\theta,\varphi,t)$, the canonical momentum is the toroidal magnetic flux $\psi_t$, the canonical coordinate is the poloidal angle $\theta$, and the canonical time the toroidal angle $\varphi$, Figure \ref{fig:flux}.  The actual clock time $t$ is just a parameter in the Hamiltonian.

\citep*{Boozer:1983} showed the topological properties of magnetic field lines at each instant in time are given by $\psi_p(\psi_t,\theta,\varphi,t)$.  The position function $\vec{x}(\psi_t,\theta,\varphi,t)$ is needed to plot the field lines in ordinary space with $\partial\vec{x}/\partial t$ giving the velocity of $(\psi_t,\theta,\varphi)$ through space, which is the field line velocity when field line topology is not changing, which implies $\psi_p$ can be taken to be independent of time.

\begin{figure}
\centerline{ \includegraphics[width=3.0 in]{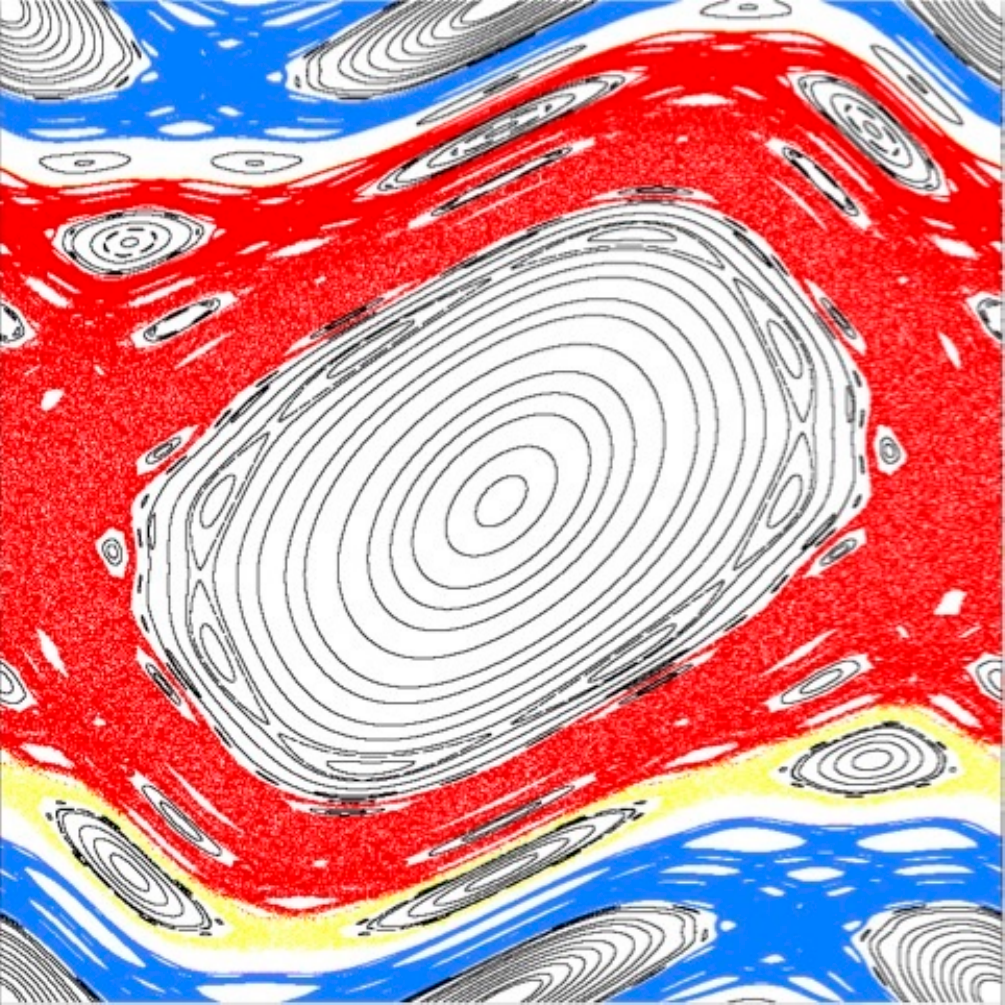}}
\caption{The Standard Map gives trajectories by iteration in a space of two periodic coordinates.  The horizontal axis is $\alpha$, the vertical axis is $\beta$, and both have a period of unity.  The $n+1$ iterate of the map  is $\alpha_{n+1} = \alpha_n + \beta_{n+1}$ and $\beta_{n+1} = \beta_n - (k/2\pi)\sin(2\pi\alpha_n)$.  Trajectories are illustrated for $k=0.975.$  Black trajectories show island chains.  The red region is a single trajectory iterated $10^6$ times, but the trajectory still escapes from that region after approximately $10^9$ iterations. This was Figure 1 in J. D. Meiss, \emph{Thirty years of turnstiles and transport}, Chaos \textbf{25}, 097602 (2015).} 
\label{fig:Stdmap}
\end{figure}

When a particular line $\vec{x}_0(\ell)$ remains within a bounded region of space, it has three possibilities as $\ell\rightarrow\infty$:  It can close on itself after a distance $\ell_0$.   It can cover a surface coming arbitrarily close to every point on a surface without ever going through the same point twice.  It can cover a volume coming arbitrarily close to every point in a volume of space without ever going through the same point twice.  All three types of trajectories are illustrated by the Standard Map, Figure \ref{fig:Stdmap}. The iteration number of the map represents the distance $\ell$.  The periodicity of both coordinates of the Standard Map makes it non-trivial to translate them into the $(\psi_t,\theta)$ coordinates of the magnetic field line Hamiltonian $\psi_p$.  \citep*{Boozer-Punjabi:2018} in their Appendix B have obtained an explicit formula for a $\psi_p(\psi_t,\theta,\varphi)$ that represents the qualitative features of stellarator equilibria including the non-resonant divertor that  \citep{Strumberger:1992} found naturally arises at the edge optimized stellarators.   Non-resonant divertors are discussed in Section \ref{Sec:NR divertor} as way to build intuitive understanding of the subtleties of cantori and turnstiles.

The behavior of neighboring magnetic field lines, which are infinitesimally separated from an arbitrary line $\vec{x}_0(\ell)$, can be studied without approximation by a particularly simple Hamiltonian \citep*{Boozer:2012}.  The study of neighboring lines, which is reviewed in Appendix \ref{Sec:separation}, showed: (1) It is possible for field lines infinitesimally separated from a given line $\vec{x}_0(\ell)$, of any of the three types, to have separations that increase or decrease exponentially with $\ell$.  There must be an equal flux of neighboring lines that diverge from and approach the line $\vec{x}_0(\ell)$ due to $\vec{\nabla}\cdot\vec{B}=0$. When lines that lie on a circle of infinitesimal radius at $\ell=0$ are followed a distance $\ell$, the circle will deform into an ellipse.  The field strength times the square of the semi-major axis of the ellipse increases as $e^{\sigma_{max}(\ell)}$, and the field strength times the square semi-minor axis decreases as $e^{-\sigma_{max}(\ell)}$. \color{black} (2) When the magnetic field depends on only two spatial coordinates, a single field line cannot cover a volume, as in the red area of Figure \ref{fig:Stdmap}.  Three spatial coordinates are required.  Two-coordinate magnetic fields, as in axisymmetry, can have X-points (more correctly X-lines) which are specific field lines from which neighboring lines exponentially approach and separate.  However, the volume in which lines have at least a certain exponentiation is itself exponentially small.  (3) Curl-free magnetic fields can exhibit field lines with an exponential separation throughout a volume.  An ideal magnetic evolution does not conserve the exponentiation $\sigma_{max}(\ell)$.  In other words, the exponentiation can be a property of the position function $\vec{x}(\psi_t,\theta,\varphi,t)$ with $\psi_p$ an unchanging function of its canonical coordinates.  The current density required for an ideal evolution to produce an exponentiation $\sigma_{max}$ is proportional to $\sigma_{max}$, Appendix \ref{Sec:j-max}.  As will be discussed in  Section \ref{Near-ideal-reconnection}, \color{black} the time required for a large scale reconnection to occur is determined by a timescale set by the ideal evolution multiplied by $\sigma_r$.  The required exponentiation  $e^{\sigma_r}$ \color{black} for large scale reconnection is the ratio of a timescale for non-ideal diffusive effects, such as resistivity, to a timescale determined by the ideal evolution.  
 

The tendency to construct two dimensional models has led to the prejudice that exponential separation of magnetic field lines throughout a volume is peculiar.   Actually any magnetic field that depends on all three spatial coordinates tends to be chaotic, which is illustrated by the design of a curl-free magnetic fields for a stellarators.  Obtaining nested magnetic surfaces, as required for a three-dimensional stellarator, rather than having the field lines fill a volume requires careful optimization.   Even a small perturbation to an optimized curl-free stellarator can destroy the surfaces.  Nevertheless, axisymmetric tokamaks are even more sensitive to external field perturbations due to their amplification by coupling to marginally stable current-driven kinks \citep{Park:2010}.

The concept of magnetic field line chaos and its definition are sufficiently subtle that Section \ref{Sec:chaos} is largely devoted to the topic.


When the magnetic field lines lie in toroidal magnetic surfaces, the poloidal angle $\theta$ can be chosen so the poloidal flux is a function of only the toroidal flux and time, $\psi_p(\psi_t,t)$.  At each instant in time, the trajectories of field lines are then given by $\psi_t$ and $\theta_0$ constant with $\theta = \theta_0+ \iota(\psi_t)\varphi$, where the rotational transform $\iota\equiv\partial\psi_p/\partial\psi_t$.  In tokamaks, it is customary to use the safety factor $q\equiv1/\iota$ instead of the rotational transform.   

 On a rational surface the rotational transform is a rational number, the ratio of two integers $\iota=N/M$.  The field lines on that surface close after $N$ poloidal transits and $M$ toroidal transits.  This identification with poloidal and toroidal transits may seem backwards from the standard convention of using $m$ as a poloidal and $n$ as a toroidal mode number as in $\cos(m\theta-n\varphi)$.  But, when $\iota=N/M$, then along a field line $\cos(m\theta-n\varphi)=\cos\Big(m\theta_0+(mN/M)\varphi-n\varphi\Big)$, which has no variation along the line since it equals $\cos(m\theta_0)$ when $m=M$ and $n=N$. \color{black}

 \citep*{Boozer:2004} showed the time derivative of the poloidal flux is the loop voltage, 
 \begin{equation}
 \frac{\partial \psi_p(\psi_t,t)}{\partial t }= V_\ell. \label{dpsip/dt}
 \end{equation} 
  At the magnetic axis, $\psi_t=0$, Equation (\ref{dpsip/dt}) is a trivial consequence of Faraday's Law plus Stokes' Theorem with the loop voltage $V_\ell$ the integral of the electric field around the axis.  The proof of this equation for an arbitrary magnetic surface is non-trivial but holds as long as the surface enclosing the toroidal flux $\psi_t$ exists.  \color{black}    On an arbitrary magnetic surface, the loop voltage is an integral along a field line at an instant of time,
\begin{equation} 
V_\ell \equiv \lim_{M_t\rightarrow\infty} \frac{1}{M_t} \int_0^{2\pi M_t} \frac{\vec{E}\cdot\vec{B}}{\vec{B}\cdot \vec{\nabla}\varphi} d\varphi \label{V_L}.
\end{equation} 
On a rational surface, each value of $\theta_0$ gives a separate closed field line, and when the loop voltage depends on $\theta_0$, the magnetic surface will split in the next instant of time to form a magnetic island.

 As proven in the paragraph below Equation (\ref{E-exp}), when $\vec{\nabla}V_\ell=0$ the topology of the magnetic field lines cannot change, and their evolution is considered ideal.  This is in contrast to ideal magnetohydrodynamics, which requires a zero resistivity $\eta$, and that implies $V_\ell=0$.   As pointed out by \citep{Boozer:2025:Constraints} and discussed in Section \ref{sec:tokamak}, the distinction between $\vec{\nabla}V_\ell=0$ and $V_\ell=0$ is important when devising strategies for avoiding disruptions in tokamaks. The condition $\vec{\nabla}V_\ell=0$ is consistent with changing the magnetic flux at the axis of a toroidal plasma without changing the magnetic field in the plasma while the condition $V_\ell=0$ is not. \color{black}

Section \ref{Sec:breakup} explains a perturbation that breaks magnetic surfaces and forms chains of magnetic islands at the rational surfaces that resonate with the rotational transform $\iota \equiv\partial \psi_p/\partial \psi_t$.  A resonant perturbation to $\psi_p$ has a toroidal mode number $n$ and a poloidal mode number $m$ with $n/m=\iota$.  The width of these islands is proportional to the square root of the resonant perturbations.  Non-resonant perturbations distort magnetic surfaces but do not break them.  Resonant perturbations that create islands cannot be ideal, they must change the form of the poloidal flux, but the position function can be held fixed.  

Non-resonant perturbations can be taken to change the position function $\vec{x}(\psi_t,\theta,\varphi,t)$ while the poloidal flux can be held fixed using a canonical transformation or what is mathematically equivalent an electric potential.  This is shown in the appendix to \citep*{Boozer:2004}.


Section \ref{Sec:chaos} discusses the subtle properties of chaotic magnetic field lines. These properties need to be understood before undertaking a study of cantori and turnstiles, which serve as boundaries on regions of chaotic magnetic field lines.

 \citep*{Chirikov:1960} gave the approximate condition for the creation of large chaotic volumes: the overlap of the islands from different resonant rational surfaces. 
 
 As discussed in Section \ref{Sec:cantori}, \color{black} the actual formation of chaotic regions is more complicated than just the overlap of island chains.  \citep*{MacKay:1984} showed that chaotic regions are formed by cantori and the turnstiles that penetrate them.  

The last magnetic surface to break between two large magnetic island chains is the most irrational surface between the resonant surfaces associated with those chains, Section \ref{Sec:cantori}.  This is quantified by the magnitude of $m$ required for the transform on this surface to satisfy $\iota(\psi_t) = n/m + \epsilon_r$ with $n$ a toroidal mode number and $m$ a poloidal mode number as $|\epsilon_r| \rightarrow0$.  As the last magnetic surface breaks, it is called a cantorus.  A  cantorus is like an irrational magnetic surface but develops holes.  These holes are places where $\vec{B}\cdot\hat{n}\neq0$ with $\hat{n}$ the unit normal to the cantorus.  The holes define highly collimated tubes of magnetic flux and must come as an inward and an outward pair due to $\vec{\nabla}\cdot\vec{B}=0$.  For this reason they are called turnstiles.  

The highly collimated flux tubes of turnstiles are responsible for the severe damage that can be caused by runaway electrons following the lines in these tubes to localized positions on the chamber walls \citep*{Boozer-Punjabi:2016}.  This collimation also responsible for the formation of a natural non-resonant divertor at the edge of a stellarator \citep*{Strumberger:1992} and \citep*{Boozer-Punjabi:2018}.  
 Section \ref{Sec:NR divertor} on non-resonant divertors provides an intuitive introduction to cantori and turntiles through an application to a practical problem. \color{black}


Section \ref{Applications} is a discussion of applications.  The first topic is magnetic reconnection. \color{black}  If the plasma evolution were perfectly ideal, a resonant perturbation would affect only the position function $\vec{x}(\psi_t,\theta,\varphi,t)$.  But, a resonant ideal perturbation  forces the magnetic field to have a delta-function current density at the resonant rational surface $\iota=n/m$.  This current causes distortions to the neighboring surfaces producing magnetic perturbations with all harmonics $k$ of the resonant perturbation $(kn)/(km) = \iota$ with $k$ any integer \citep{Huang-Hudson:2022}.  In a torus, the toroidicity produces magnetic perturbations with poloidal harmonics of strength $\epsilon^k$ at $m\pm k$ with $\epsilon$ the inverse aspect ratio of the torus.  Any rational number can be produced by the harmonics of $n/m$ together with the harmonics of $m$ produced by toroidicity.  The more ideal the perturbation, the more contorted the magnetic surfaces become, which leads to their rapid breaking by an arbitrarily small non-ideal effect, such as resistivity \citep*{Boozer:2022}.   Section \ref{Applications} also discusses applications to tokamak disruptions and to stellarators.

Section \ref{Sec:Discussion} is an overview of the paper.  The Appendix reviews the behavior of magnetic field lines that are infinitesimally separated from an arbitrary given line, $\vec{x}_0(\ell)$. \color{black}


\section{Representation of a magnetic field and its field lines  } \label{Sec:B-rep}

 \citep*{Boozer:1983} showed that the magnetic field in a toroidal region can always be written as
\begin{eqnarray}
2\pi\vec{B} = \vec{\nabla}\psi_t \times \vec{\nabla}\theta + \vec{\nabla}\varphi \times \vec{\nabla} \psi_p,  \label{Gen-B}
\end{eqnarray}
where $\theta$ and $\varphi$ are arbitrary poloidal angles, Figure \ref{fig:flux}.   The toroidal flux $\psi_t$ is the toroidal flux enclosed by a constant-$\psi_t$ surface, and the poloidal flux $\psi_p$ is the flux penetrating the central hole in the torus formed by a constant-$\psi_p$ surface.

Equation (\ref{Gen-B}) follows from writing the vector potential $\vec{A}$, where $\vec{B}=\vec{\nabla}\times\vec{A}$, using three coordinates $(r,\theta,\varphi)$.  Then, $\vec{A}= A_r\vec{\nabla}r +A_\theta\vec{\nabla}\theta +A_r\vec{\nabla}\varphi +\vec{\nabla}g$.  The gauge can be chosen so $\partial g(r,\theta,\varphi)/\partial r=-A_r$.  Let $\psi_t/2\pi \equiv A_\theta + \partial g/\partial\theta$, and $\psi_p/2\pi \equiv -(A_\varphi + \partial g/\partial\varphi)$, so $2\pi \vec{A} = \psi_t \vec{\nabla}\theta - \psi_p \vec{\nabla} \varphi$, which has Equation (\ref{Gen-B}) as its curl.

In general the poloidal flux depends on all three spatial coordinates and time, $\psi_p(\psi_t,\theta,\varphi,t)$.  \citep*{Boozer:1983} showed the trajectories of the magnetic field lines are given by a Hamiltonian $\psi_p(\psi_t,\theta,\varphi,t)$ in canonical form with time $t$ just a parameter:
\begin{eqnarray}
\frac{d\psi_t}{d\varphi} &\equiv& \frac{\vec{B}\cdot\vec{\nabla}\psi_t}{\vec{B}\cdot\vec{\nabla}\varphi}   \\
&=& - \frac{\partial \psi_p(\psi_t,\theta,\varphi,t)}{\partial\theta} \hspace{0.2in} \mbox{and   } \label{dpsi_t}\\
\frac{d\theta}{d\varphi} &\equiv& \frac{\vec{B}\cdot\vec{\nabla}\theta}{\vec{B}\cdot\vec{\nabla}\varphi} \\
&=& \frac{\partial \psi_p(\psi_t,\theta,\varphi,t)}{\partial\psi_t}. \label{dtheta}
\end{eqnarray}

These equations imply that the variation of $\psi_p$ along a trajectory $d\psi_p/d\varphi =\partial \psi_p/\partial\varphi$ since the terms involving the $\varphi$ and $\theta$ derivatives cancel.  When $\partial\psi_p/\partial\varphi=0$, the poloidal angle $\theta$ can be chosen so the toroidal and poloidal fluxes are functions each other, $\psi_p(\psi_t)$.  This choice of the poloidal angle gives what are called magnetic coordinates.  In a symmetric torus, magnetic coordinates always exist, and they may exist in non-axisymmetric situations, as in stellarators, but careful design is required.

In an axisymmetric torus, the magnetic surfaces can be bounded by X-points, as is well known because of tokamak divertors.  At an X-point, the gradient of the magnetic poloidal angle vanishes, $\vec{\nabla}\theta=0$, which implies $\vec{B}\cdot\vec{\nabla}\theta=0$ and requires $\vec{\nabla}\psi_t\rightarrow\infty$ since $(\vec{\nabla}\psi_t \times \vec{\nabla}\theta)\cdot\vec{\nabla}\varphi = 2\pi \vec{B}\cdot\vec{\nabla}\varphi \neq 0$.  

The distance between neighboring irrational magnetic surfaces $\Delta(\theta,\varphi)\equiv |\psi_t|/|\vec{\nabla}\psi_t|$ can have an exponentially large ratio between its maximum and minimun, $\Delta_{max}/\Delta_{min}=e^{2\sigma}$, despite all the surfaces remaining perfect.  As discussed in Section \ref{Sec:reconnection}, when $\sigma \gtrsim 20$ even the small resistivity in fusion plasmas will cause the magnetic surfaces to break and form a region of volume-filling chaotic magnetic field lines.  Since $(\vec{\nabla}\psi_t \times \vec{\nabla}\theta)\cdot\vec{\nabla}\varphi = 2\pi \vec{B}\cdot\vec{\nabla}\varphi$ varies little over a magnetic surface compared to $e^{20}$, places where neighboring surfaces are close together must be places where the field lines in the surfaces are far apart and vice versa  \citep*{Boozer:2022}.



\section{Breakup of magnetic surfaces  } \label{Sec:breakup}



Magnetic surfaces breakup by the formation of magnetic islands when a rational surface $\iota = n/m$ is perturbed with the width of the island proportional to the square root of the perturbation.  The interior of an island has magnetic surfaces.  The magnetic surfaces in a curl-free stellarator field can be considered to be those of an island.  Figure \ref{fig:Stdmap} shows islands can have islands in their interiors.

A small perturbation to the poloidal flux, $\psi_p = \psi_p^{(0)}(\psi) +  \psi_p^{(1)}(\psi)\sin(m\theta-n\varphi)$, is equivalent to a small field perpendicular to the unperturbed magnetic surfaces $\vec{B}_1\cdot\vec{\nabla}\psi_t^{(0)} = - (\partial \psi_p^{(1)}/\partial\theta)(\vec{B}_0\cdot\vec{\nabla}\varphi) $.  The distortion to the magnetic surfaces is given by Equation (\ref{dpsi_t}), which is equivalent to the perturbation to the toroidal flux $\psi_t^{(1)}$ obeying $\vec{B}_0\cdot\vec{\nabla}\psi_t^{(1)} + \vec{B}_1\cdot\vec{\nabla}\psi_t^{(0)} =0$.  Let $\theta_h\equiv m\theta-n\varphi$, then $\vec{B}_0\cdot\vec{\nabla}\theta_h = (m\iota(\psi_t)-n)\vec{B}\cdot\vec{\nabla}\varphi$.   Near $\iota(\psi_t) = n/m$, the result of the perturbation field appears to be singular, but this resonant surface problem can be resolved by writing $\iota(\psi_t) = \iota(\psi_t^{(0)}) + \iota' \psi_t^{(1)} + \cdots$, where $\iota'\equiv d\iota/d\psi_t$.  Sufficiently close to the resonant surface, $\iota(\psi_t) = n/m$ and for a sufficiently small perturbation
\begin{eqnarray}
\iota' \psi_t^{(1)} \frac{\partial \psi_t^{(1)} }{\partial\theta_h} &=& - \frac{ \vec{B}\cdot\vec{\nabla}\psi_t^{(0)} }{\vec{B}_0\cdot\vec{\nabla}\varphi} \label{island-diff}\\
&=& m\psi_p^{(1)}\cos\theta_h \\
\psi_t^{(1)} &=& \pm \sqrt{ \frac{4\psi_p^{(1)}}{|m\iota'|}\left(s^2 - \sin^2(\theta_h/2)\right)}, \hspace{0.3in}
\end{eqnarray}
where $s^2$ is the additive constant from Equation (\ref{island-diff}).  The identity $\cos x = 1-2\sin^2 (x/2)$ was used.

An island chain forms at every rational surface that resonates with the perturbation.  Realistic magnetic perturbations have many Fourier harmonics.  Only in a cylinder with two symmetry directions is a single Fourier harmonic perturbation possible.  \citep*{Chirikov:1960} noted that when island chains from different rational surfaces overlap the region between the rational surfaces becomes chaotic.   


%


\section{Chaos \label{Sec:chaos}}

The field lines of chaotic magnetic fields are generally thought to have two properties: (1) Each field line has neighboring lines that have a separation that depends exponentially on the distance $\ell$ along the line throughout a volume.  (2) A single field line covers the whole chaotic volume as in the red region of Figure \ref{fig:Stdmap}.  

Remarkably, the first property does not imply the second, and it is the first property that is of central importance in reconnection theory.  It is possible for a magnetic field lines that cover surfaces, not volumes, to exponentially separate from neighboring lines throughout a volume, which is indeed what happens as the plasma response approaches an ideal response $|\vec{\nabla}V_\ell|=0$.  The possibility of field line chaos in the sense of an arbitrarily large exponentiation in separation from neighboring lines  is shown in Appendix \ref{Sec:sep} and discussed in the Introduction.  

\begin{figure}
\centerline{ \includegraphics[width=3.2 in]{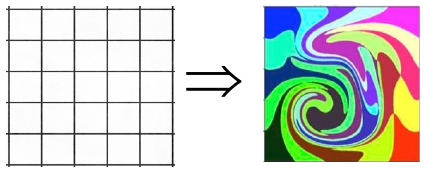}}
\caption{ A magnetic field $\vec{B}(\vec{x},t)$ can be thought of as consisting of tubes of magnetic flux by placing a gridded surface across the field.  Each tube is defined by the magnetic field lines that pass through the perimeters of the grid cells.  When the field is chaotic, the perimeter of each cell becomes exponentially longer when the grid is replotted after each line on the perimeters is followed for a distance $\ell$.   But, each cell contains exactly the same field lines and has precisely the same neighboring cells.  When the magnetic field is evolving ideally with a chaotic velocity $\vec{u}_\bot$, a similar distortion of the grid occurs when the grid is replotted using the location of each line on the perimeters after a time $t$.  The figure shows the distortion of a $5\times5$ array.   This is Figure 1 of Boozer, Phys. Plasmas \textbf{32}, 052106 (2025).  The distorted grid is part of Figure 5 of Y.-M. Huang and A. Bhattacharjee, Phys. Plasmas 29, 122902 (2022), which was based on a chaotic evolution defined by A. H. Boozer and T. Elder, Phys. Plasmas \textbf{28}, 062303 (2021).  Boozer and Elder illustrated distortions of ideally evolving flux tubes up to a factor $\sim 10^7$.  } 
\label{fig:tubes}
\end{figure}

Chaos in the sense of exponentiation allows a magnetic field $\vec{B}(\vec{x})$ that has variations only on long spatial scales to have structures in the magnetic field lines on arbitrarily small spatial scales.  As illustrated in Figure \ref{fig:tubes}, the exponentiation in separation of neighboring lines causes an exponentially large distortion in any tube of magnetic flux including the tubes formed by neighboring irrational magnetic surfaces.  This distortion is of fundamental importance to the physics of magnetic reconnection, Section \ref{Sec:reconnection}, though generally ignored in the reconnection literature.   Neither localized currents nor localized magnetic fields are required.  Chaos is observed in curl-free magnetic fields that depend on all three spatial coordinates and are not carefully designed to avoid chaos.  Nevertheless, a large exponentiation while field lines lie on perfect magnetic surfaces does not seem possible in a curl-free field.

Chaotic magnetic field lines in which a single field line covers the whole chaotic volume also separate exponentially when followed far enough in $\ell$.  But, the existence of cantori with highly collimated turnstiles makes the implication of the first property of chaotic field lines by the second also subtle.  As discussed by \citep*{MacKay:1984} and in the Introduction, the formation of chaotic regions in which a single line covers a volume involves cantori, which can appear to be just an irrational magnetic surface for an arbitrarily long distance $\ell$ along one of the field lines in the surface.  These subtleties are also discussed in Section \ref{Sec:cantori} and must be understood to determine the relevance of magnetohydrodynamic (MHD) simulations to actual plasma behavior.

A subtlety of a field line filling a volume is illustrated by a point made by  \citep*{Meiss:2015}.   Although the red area in Figure \ref{fig:Stdmap} was formed by a single chaotic trajectory during a million iterations, that trajectory escaped from the red area after approximately a billion iterations.   The implication is that the boundary of the red area was a cantorus penetrated by turnstiles that occupy only a tiny fraction of the area of the boundary.  \citep*{Punjabi:2025} found that outermost confining magnetic surface of a stellarator has a similar subtlety.  A region with a width of a few per cent of the $\psi_t$ enclosed by the outermost confining surface confines magnetic field lines while they make transits through hundreds of thousands of stellarator periods, but they eventually strike the wall.  There is presumably little difference in the behavior of the plasma on these lines than that on lines that have perfect confinement.  

The Brouwer fixed-point theorem of topology guarantees that field lines that close on themselves exist within any bounded chaotic  field-line region.  These lines can still be chaotic in having the first property of neighboring field lines with an exponential separation although obviously not the second property of covering the entire chaotic region.  Although these closed lines can have neighboring lines that exponentially separate, this is not necessary because they can be the magnetic axis of a magnetic island.  Figure \ref{fig:Stdmap} illustrates the existence of such islands within the red region. 

When the electron pressure, $p_e$, has a gradient along chaotic magnetic field lines,  \citep{Boozer:2024:Electric} showed that the electric field required for quasi-neutrality gives $\vec{E}\times\vec{B}$ flows that lead to plasma transport at a Bohm-like rate.   A variation in the electric potential, $en\partial \Phi/\partial \ell =  \partial p_e/\partial \ell$, where $n$ is the electron number density, is required to keep the rapidly moving electrons together with the slowly moving ions.   The exponentially large dependence of the separation of neighboring field lines with $\ell$ leads to increasing  $\vec{E}\times\vec{B}$ flows until a balance in electron and ion transport is achieved.  The required Bohm-like level of transport across the magnetic field lines is important for sweeping in impurities in tokamak disruptions and for transport in non-resonant stellarator divertors.

The velocity of a flow can also be chaotic by having chaotic streamlines.  In a chaotic flow, each streamline, $\vec{x}(t)$ with $d\vec{x}/dt = \vec{u}(\vec{x},t)$, has neighboring streamlines that increase their separation exponentially with time.   A chaotic magnetic field requires a dependence on all three spatial coordinates.  A chaotic magnetic field has chaotic magnetic field lines, which are defined at fixed instants in time.  However, a chaotic flow requires a dependence on only two spatial coordinates when the flow is time dependent.  The failure to appreciate the implications explains the origins of fundamental misconceptions in the theory of magnetic reconnection, Section \ref{Sec:reconnection}.   As discussed in \citep{Boozer:2025:Magnetic}, when the flow velocity of magnetic field lines is chaotic, the ideal evolution of a magnetic field in a system with only two spatial coordinates produces an exponential increase of the field strength.  When the field depends on all three spatial coordinates, a chaotic flow velocity makes the field lines chaotic while producing only a moderate increase in the magnetic field strength.


\section{Cantori and turnstiles \label{Sec:cantori}}

In an evolving $\vec{B}$, the last magnetic surface to break is the surface with iota most difficult to approximate by a rational.  As an irrational surface breaks, it forms a cantorus.   Cantori can also exist in static magnetic fields in the region in which confining magnetic surfaces transition into open magnetic field lines that strike the walls, as in a non-resonant stellarator divertor.

A cantorus has places at which $\vec{B}\cdot\hat{n}\neq0$, where $\hat{n}$ is the normal to the surface.  These places act as holes through which magnetic field lines can pass in what would otherwise be an irrational magnetic surface.  These holes can be arbitrarily small, which implies the flux that passes through the holes forms highly collimated magnetic flux tubes called turnstiles.  $\vec{\nabla}\cdot\vec{B}=0$ implies turnstiles must occur in pairs, one carrying flux inwards and the other outwards.


\subsection{Efficient way to obtain information about cantori and turnstiles}   \color{black}

Conventionally subtleties of magnetic field line evolution are understood using Poincar\'e plots.   \citep*{Boozer:2025:Efficient} has shown that far more information can be obtained by Fourier decomposing $R(\varphi)$ and $Z(\varphi)$ of magnetic field lines that are followed in ordinary $(R,\varphi,Z)$ cylindrical coordinates to find surfaces on which magnetic field lines lie or near which they linger.  Knowing these surfaces, $\vec{B}\cdot\hat{n}$ can be determined, cantori identified, the magnitude and collumation of the magnetic flux that leaks through these surfaces in turnstiles can be found.

On a magnetic surface, the magnetic poloidal angle obeys $\Theta=\iota_s \varphi$, where $\iota_s$ is a constant, the rotational transform.  The expansion of $Z(\Theta,\varphi)=\sum_{mn} Z_{mn}\cos(m\Theta-n\varphi)$ can be found using a Gaussian function of width $w$, which satisfies $\int_0^\infty G(\varphi) \cos(\Omega\varphi) d\varphi= \exp(-(w \Omega)^2/2)$.  The function $Z_f(\omega) \equiv 2 \int_0^\infty G(\varphi) \cos(\omega\zeta) Z(\varphi) d\varphi$ can be found by a Fast Fourier Transform.  When $w|\omega_{mn}| >> 1$, where $\omega_{mn}\equiv m\iota_s -n$, the function $Z_f(\omega)$ has the form $\sum_{mn} Z_{mn}\exp(-(w^2/2)(\omega-\omega_{mn})^2)$, which determines the Fourier coefficients and $\iota_s$.   For simplicity, it has been assumed only sine or cosine terms are required for the Fourier decomposition for $R$ and $Z$.

When the field line is chaotic, the Gaussian peaks have an additional standard deviation due to their chaotic spreading. When the chaos is small, the standard deviation decreases with the width  $w$ of the Gaussian window function. Whether the field line is chaotic or not is determined by making the Gaussian width  $w$ greater, appropriately extending the integration range, and observing whether the Gaussian peaks become narrower or broader.

The position vector
\begin{eqnarray}
\vec{x}(\psi_t,\Theta,\varphi) &=& \hat{R}(\varphi) \sum R_{mn}(\psi_t) \sin(m\Theta-n\varphi) \nonumber\\ &+& \hat{Z} \sum Z_{mn}(\psi_t)\cos(m\Theta-n\varphi).
\end{eqnarray}
The inverse of the Jacobian, $(\vec{\nabla}\psi_t \times \vec{\nabla}\Theta)\cdot \vec{\nabla}\varphi = 2\pi \vec{B} \cdot\vec{\nabla}\varphi$, and the toroidal field $B_t = R\vec{B} \cdot\vec{\nabla}\varphi$ is almost constant on a surface.  

The distance between surfaces
\begin{eqnarray}
\Delta &=& \left|\frac{\psi_t}{\vec{\nabla}\psi_t}\right| \\
&=& \frac{R \psi_t}{2\pi B_t} \frac{1}{\Big| \frac{\partial \vec{x}}{\partial\Theta}  \times \frac{\partial \vec{x}}{\partial\varphi} \Big|}.\\
\sigma_s(\psi_t)& \equiv & \ln\left(\sqrt{\frac{\Delta_{max}}{\Delta_{min}}} \right).
\end{eqnarray}
Reconnection with neighboring surfaces becomes exponentially easier the larger the ratio of the maximum to the minimum separation on the surface.   The larger $\sigma_s$ is the more localized is reconnection to points on the surface where the minimum separation $\Delta_{min}$ is located.  

One subtlety is that there is no guarantee that surfaces do not overlap.  When there is no chaos, so the widths of the Gaussian lines decrease as $w$ increases, ensuring surfaces do not overlap should be a matter of numerical resolution.  Nevertheless, fast disruptions in high temperature tokamaks could in principle involve ratios of $\Delta_{max}/\Delta_{min}\sim 10^{10}$, which seems daunting.  Of course far smaller ratios should clearly identify the points on the surface at which reconnection takes place and turnstiles form.  The actual locations of turnstiles are the places at which $\vec{B}\cdot\hat{n}\neq0$, where the unit normal to the surface is
\begin{eqnarray}
\hat{n} &\equiv&  \frac{\frac{\partial \vec{x}}{\partial\Theta}  \times \frac{\partial \vec{x}}{\partial\varphi}}{\Big| \frac{\partial \vec{x}}{\partial\Theta}  \times \frac{\partial \vec{x}}{\partial\varphi} \Big|}
\end{eqnarray}
It is important that studies be done to obtain the minimum width of turnstiles that can be determined in this way.


\subsection{Subtleties of cantori and turnstiles in non-resonanant divertors \label{Sec:NR divertor} }

An intuitive understanding of subtleties of cantori and turnstiles can be obtained from a study of the localization of  the plasma flows in a stellarator non-resonant divertor.   This can be done by following field lines from the point where they enter the plasma chamber until they leave.  The strong helical fields outside of the magnetic surfaces of a stellarator cause almost all magnetic field lines to exit the plasma chamber within approximately a stellarator period of where they entered.  However, there are narrow helical stripes on the walls, which take up only a few percent of the wall area, with entry points that give lines that make many toroidal transits before restriking the wall.  Indeed, there are entry points within these stripes from which field lines make an arbitrarily large number of toroidal transits \citep{Punjabi:2025}.  The field lines that come closest to the confined plasma take the largest number of toroidal transits between wall entry and departure.  These lines form a tube of magnetic flux that can be filled with plasma by cross-field diffusion and produce the divertor.  This  flux tube is roughly circular but strikes the wall in a helical stripe since the magnetic field component tangential to the stellarator walls is strong in comparison to the normal field.  

The behavior of field lines in non-resonant divertors is understandable due the magnetic field just outside the last confining magnetic having a number of important cantori with the cantori closer to the outermost confining magnetic surface having narrower turnstiles.  Once a field line passes through the outermost important turnstile, which occupies only a few percent of the cantorus area, it takes many toroidal transits to find an exit hole which must have the same small magnetic flux as the entry turnnstile.  Once the field line exits that turnstile, the line strikes the wall in a fraction of a stellarator period, which is too short a distance for the exponentiation of the chaos to destroy the collimation produced by the small hole defined by the turnstile.  

While the field line is making toroidal transits between the entry and the exit turnstile of the outermost important cantorus, it can intercept an even smaller turnstile of a cantoris that is closer to the last confining magnetic surface.  These lines become trapped for an even  larger number of toroidal transits before they intercept an existing turnstile.  The strong correlation between the closeness with which a field line approaches the outermost confining magnetic surface and the number of toroidal transits the line makes is illustrated in \citep{Garcia:2025} and \citep{Punjabi:2025}.  

The location of the outermost important turnstile is determined by a strong curvature in the outermost magnetic surface, which are called helical edges.  They resemble the X-lines of a tokamak divertor.  Unlike X-lines, which are toroidally continuous in axisymmetry and require the rotational transform to be zero, helical edges extend only over a fraction of stellarator period.  A sharp helical edge on a magnetic surface, or an X-line of a tokamak divertor, can only be produced by currents far more distant than the radius of curvature of the magnetic surface when the edge is parallel to a magnetic field line.  Since the helical edge is not a closed curve, it does not define the rotational transform of the surface.  \citep{Bader:2018} has shown the locations on the wall that are struck by the field lines of a non-resonant divertor are robust against plasma changes and are determined by strong variation in the curvature of the outermost magnetic surface.


\color{red} \color{black}

When a clear separatrix defines the transition from the confinement region to a divertor, as in either a tokamak or in a W7-X island divertor, there is no resilience of the strike point locations to changes, for example to the edge rotational transform.   The outgoing field lines beyond the separatrix carry the plasma flow towards the divertor chamber.  As discussed by \citep{Garcia:2025}, resilience exists only when there are no such field lines.  \citep{Boozer:2024:Stellarators} has pointed out that compared to divertors defined by a separatrix, non-resonant divertors can have three basic advantages: (1) Resilience, the location at which the diverted plasma reaches the walls is insensitive to plasma conditions.  (2) The width of the region at the plasma edge that flows into the divertor has a controllable width.  It can be made sufficiently broad to avoid neutrals impending on the plasma from charge exchanging with high energy plasma ions, which would result in high energy neutrals that erode the walls.  (3) The controllable width and, therefore, confinement properties of non-resonant divertors allow high-Z impurities to be added.  The shortness of their confinement time limits their diffusion into the plasma interior.  High-Z impurity radiation at the plasma edge seems to be the only way to adequately spread the out coming power over the walls.
\color{black}


\section{Applications \label{Applications}}

Once the concepts of magnetic field line chaos, cantori, and turnstiles are understood their importance to determining the behavior of plasmas becomes clear.  This section discusses three araes of application: magnetic reconnection, tokamak disruptios, and stellarator physics.
\color{black}

\subsection{Magnetic reconnection \label{Sec:reconnection}}

Three concepts are needed to understand magnetic reconnection: (1) magnetic field line chaos, (2) the magnetic field line velocity, $\vec{u}_\bot$, and (3) and its chaos. The magnetic field line velocity $\vec{u}_\bot$ is distinct from the plasma velocity $\vec{v}$. 


\subsubsection{Magnetic field line velocity $\vec{u}_\bot$}

The magnetic field line velocity  $\vec{u}_\bot$ is defined by the purely mathematical equation 
\begin{equation}
\vec{E}+\vec{u}_\bot \times \vec{B} = - \vec{\nabla}\Phi + V_\ell \vec{\nabla}\frac{\varphi}{2\pi}. \label{E-exp}
\end{equation}

The loop voltage in a region of spatially bounded field lines is defined by Equation (\ref{V_L}).  When $\vec{\nabla} V_\ell=0$ the magnetic evolution is ideal. Faraday's law then implies the evolution of Clebsch coordinates $\alpha$ and $\beta$ where $\vec{B}=\vec{\nabla}\alpha\times\vec{\nabla}\beta$ is consistent with $d\alpha/dt = \partial \alpha/\partial t + \vec{u}_\bot\cdot\vec{\nabla}\alpha =0$ and $d\beta/dt = \partial \beta/\partial t + \vec{u}_\bot\cdot\vec{\nabla}\beta=0$.  Since $\alpha$ and $\beta$ define a magnetic field line, $\vec{u}_\bot$ is the velocity of the magnetic field lines when the gradient of the loop voltage is zero.  This is in essence the proof given in \citep*{Newcomb:1958}.  The physical interpretation of $\vec{u}_\bot$ is subtle when $\vec{\nabla} V_\ell \neq 0$, but $\vec{u}_\bot$ is still defined and for simplicity of language can be called the magnetic field line velocity.  The fact that $\vec{u}_\bot$ depends on the frame of reference that is used, which can change if canonical transformations are made to $(\psi_t,\theta,\varphi)$ as the system evolves, has created confusion about the existence of $\vec{u}_\bot$.

Equation (\ref{E-exp}) for $\vec{E}$ follows from mathematics alone---the representation of vectors in a three dimensional space.  Although \citep*{Boozer:2025:Magnetic} has shown that Equation (\ref{E-exp}) is far more generally valid, the proof is more obvious when $\vec{B}\cdot\vec{\nabla}\varphi\neq0$.  An arbitrary vector $\vec{E}$ in three-space can be represented in terms of another arbitrary vector $\vec{B}$ when all three components of $\vec{E}$ can be represented.  The component $\vec{B}\cdot\vec{E} = -\vec{B}\cdot\vec{\nabla}\Phi + V_\ell \vec{B}\cdot\vec{\nabla}\varphi$ is equivalent to
\begin{equation}
\left(\frac{\partial \Phi}{\partial \varphi} \right)_{\alpha \beta} = - \frac{\vec{B}\cdot\vec{E}}{\vec{B}\cdot\vec{\nabla}\varphi} + \frac{V_\ell}{2\pi}.
\end{equation}
Locally this equation for $\Phi$ along a magnetic field line always has a solution with $V_{\ell}=0$.  However, a loop voltage $V_\ell$ is generally required to make $\Phi$ a periodic function $\varphi$ in a torus or when there are two boundary conditions on $\Phi$ to be satisfied as there generally are when field lines strike walls.  The velocity $\vec{u}_\bot$ can always be chosen so the two components of $\vec{E}$ perpendicular to $\vec{B}$ are represented.

Faraday's Law and Equation (\ref{E-exp}) imply that without approximation
\begin{eqnarray}
\frac{\partial \vec{B}}{\partial t} = \vec{\nabla} \times (\vec{u}_\bot \times \vec{B}) + \frac{\vec{\nabla}V_\ell \times \vec{\nabla}\varphi }{2\pi}. \label{Ad-diff}
\end{eqnarray}
When the term $(\vec{\nabla}V_\ell \times \vec{\nabla}\varphi)/2\pi$ involves second derivatives of $\vec{B}$ with respect to position then it is diffusive.  This is the case when there is a term $\eta\vec{j}$ in the expression for the electric field.  Equation (\ref{Ad-diff}) is then an advection diffusion equation. 

\citep*{Aref:1984} illustrated the exponentially large effect on mixing in fluids when the velocity with which they are stirred is chaotic.  \citep*{Tang:1999} used Lagrangian coordinates to show the effective spatial diffusion increases exponentially in time.


\subsubsection{Magnetic reconnection when $\vec{B}$ is nearly ideal \label{Near-ideal-reconnection} }

When  $\vec{B}$ is ideal and $\vec{u}_\bot$ is chaotic, the surfaces that enclose a volume of magnetic flux increase their surface area exponentially without limit even though the enclosed magnetic flux is a constant. The exponential increase in the flux tube perimeters in Figure \ref{fig:tubes} correspond to an exponential increase in surface area if a three rather than a two dimensional plot were made.  \color{black}  (Boozer and Elder 2021) gave an example of smooth large-scale non-turbulent magnetic field line flow $\vec{u}_\bot$ that is chaotic.  Their example is particularly interesting because it puts no helicity into the system.   As they discuss, unlike magnetic energy, the dissipation of magnetic helicity cannot be significantly enhanced by turbulence.  In a system like the solar corona in which magnetic flux tubes are driven by plasma motion in the photosphere, helicity buildup cannot be bounded other than by the ejection of the loop from the sun, a coronal mass ejection.


As contortions of the enclosing surfaces become exponentially greater, two surfaces that were initially well separated will develop points at which their separation decreases exponentially.   These points and the timescale for their development are determined by the ideal evolution of $\vec{B}$ when non-ideal effects are extremely small, $\left|\vec{\nabla}V_{\ell}\right|\rightarrow0$.

When initially well-separated flux tubes have points where their separations decrease exponentially in time due to the ideal evolution, resistive diffusion $\eta/\mu_0$ will eventually interdiffuse lines between the tubes---for arbitrarily small $\eta/\mu_0$; $\tau_{rec} \approx \tau_u\ln(\frac{ \tau_\eta}{\tau_u })$. Streamlines of $\vec{u}_\bot$ e-fold on the timescale $\tau_u$, and $\tau_\eta\equiv\mu_0 a^2/\eta$ is the resistive timescale.  The property of logarithms implies $\tau_{rec} \approx 20\tau_u$ even when $\tau_\eta/\tau_u=10^{10}$.


\subsubsection{Limitation on reconnection due to electron inertia \label{sec:e-inertia}}

The finite electron mass can cause magnetic reconnection even in the absence of any other non-ideal effect.  However, \citep*{Boozer:2026:Electron} \color{black} proved a remarkable result.   When electron inertia is the only non-ideal effect in the evolution of a magnetic field $\vec{B}$, there is a related field that evolves ideally.  This field is $\vec{\mathcal{B}} \equiv \vec{B} + \vec{\nabla}\times \left( (c/\omega_{pe})^2\mu_0\vec{j} \right)$ with $\omega_{pe}$ the plasma frequency  and $\vec{j}$ the current density.   In three dimensional space, the practical importance of the ideal evolution of $\vec{\mathcal{B}}$ on its reconnection appears limited.  Section \ref{Near-ideal-reconnection} shows that when the evolution velocity of modified field is chaotic, $\vec{\mathcal{B}}$ will reconnect on a timescale that depends only logarithmically on any non-ideal effect that is diffusive, such as resistivity.  



\subsubsection{Magnetic surface breakup and electrostatic  instabilities \label{ITG-chaos}}

\citep{Nevins:2011},  \citep*{Connor:2013}, and \citep{Terry:2015} have discussed magnetic surface breakup that is produced by what are called electrostatic instabilities, primarily the ion temperature gradient (ITG) instabilities.  The focus has been on the enhancement of the electron heat transport, which is not large when the plasma $\beta\equiv 2\mu_0p/B^2$ is small.  

An effect that can arise before the electron heat transport is appreciable is a modification of the radial electric field required to preserve quasi-neutrality in stellarators.  This modification could be of great practical importance because neoclassical transport in stellarators usually gives more rapid ion than electron transport, which requires an electric field that confines ions---particularly high charge state impurities.  \citep*{Helander:2008} discussed the lack of effect of electrostatic microturbulence on the large scale radial electric field though the turbulent Reynolds stress can produce zonal flows that tend reduce the microturbulent transport. They did not discuss the effect of the breaking of the magnetic surfaces by the turbulence.  The enhanced radial transport of electrons due to broken surfaces does modify the radial electric field, but the required momentum transport to break the surfaces is subtle.  \citep{Boozer:2026:Non-Ambipolarity} \color{black} has shown the radial transport of the electrons gives an effective perpendicular viscosity force.  \citep{Alcison:2023} found a large expulsion of impurities in W7-X experiments when the microturbulent transport exceeded the expected neoclassical, which requires an explanation.

A slab-model in which a constant magnetic field $B_z\hat{z}$ is subjected to a perturbation $\tilde{B}_x\hat{x}$ illustrates the relationship between the magnetic field line velocity $\tilde{u}_x\hat{x}$ and the plasma velocity $\tilde{v}_x\hat{x} + \tilde{v}_z\hat{z}$.  Using Equation (\ref{Ad-diff}) with an ideal perturbation, $\vec{\nabla}V_\ell=0$, Faraday's Law implies $ \partial_t \tilde{B}_x = B_z\partial_z \tilde{u}_x$, so $ \partial_z \partial_t \tilde{B}_x =  B_z\partial_z^2 \tilde{u}_x$.  Ampere's Law implies $ \partial_z \tilde{B}_x = \mu_0 j_y$ and force balance gives $j_y B_z = m_i n \partial_t \tilde{v}_x$, which leads to the equation $ \partial_z \partial_t \tilde{B}_x =  \frac{\mu_0 m_i n}{B_z}  \partial_t^2 \tilde{v}_x.$   Equating the two expressions for $ \partial_z \partial_t \tilde{B}_x$,
 \begin{eqnarray}
  \frac{\partial^2 \tilde{u}_x}{\partial z^2} &=&  \frac{\mu_0 m_i n}{B_z^2} \frac{\partial^2 \tilde{v}_x}{\partial t^2}\\
 &=&  \frac{1}{V_A^2} \frac{\partial^2 \tilde{v}_x}{\partial t^2}.
 \end{eqnarray}
  When the phase velocity of $\tilde{v}_x$ along $z$ equals $V_A$, the Alfv\'en speed, $\tilde{u}_x=\tilde{v}_x$, and the whole motion of the plasma is due to the motion of the magnetic field lines, which means an Alfv\'en wave.  When the phase velocity along $z$ is approximately the ion thermal speed $V_i$, then $\tilde{u}_x = (V_i^2/V_A^2)\tilde{v}_x$.  The ratio $V_i^2/V_A^2 \sim \beta$, and the magnetic field lines move little compared to the plasma motion.  Nevertheless, the velocity of the magnetic field lines is turbulent and, therefore, chaotic, which implies magnetic reconnection that breaks the magnetic surfaces will quickly occur, regardless of how small non-ideal effects represented by the loop voltage may be.



\subsection{Tokamak disruptions \label{sec:tokamak}}

As noted in the Introduction, the current density required to create an exponentiation $\sigma$ is only linearly dependent of $\sigma$, see Appendix \ref{Sec:j-max}, as is the timescale for large scale reconnection, see Section \ref{Sec:reconnection}.  Section \ref{Sec:chaos} showed an ideal perturbation can lead to chaotic field lines through a large volume of the plasma even though perfect magnetic surfaces are preserved.  Section \ref{Sec:reconnection} shows that time required before reconnection occurs is essentially determined by timescale on which the exponentiation of the chaos grows.  The time required for a large scale reconnection has only  a logarithmic dependence on resistivity, which makes this time essentially independent of resistivity when the resistivity is small.  Shortly after the paper \citep*{Boozer:2022} was was published showing this, \citep{Jardin:2022} carried out simulations of disruptions that were extremely fast due to an ideal mode becoming unstable.  Their simulations confirmed that the time to disruption was essentially independent of the plasma resistivity to the extent the code could maintain adequate resolution.  That is, any large scale mode, resistive or ideal, can produce a disruption on a timescale determined by  the growth rate of the mode unless it self-stabilizes at a sufficiently small amplitude.  Non-self-stabilizing modes could cause a stellarator to disrupt, but such modes are easier to avoid in stellarators than in tokamaks.

When the magnetic evolution is near ideal, as in a high-temperature tokamak,  \citep*{Boozer:2025:Magnetic} showed only a small fraction of the energy released by the reconnection is directly dissipated---most goes into Alfv\'en waves.  This is unlike  reconnection in two-coordinate systems in which the energy released by the reconnection is damped on the same timescale as the reconnection.  \citep*{Boozer:2020} showed that shear Alfv\'en waves moving along the reconnecting field lines would cause the flattening of the $j_{||}/B$ profile, which leads to the current spike observed in disruptions.  The simulations of \citep{Nardon:2023} were consistent with the predicted timescale, and they saw magnetic fluctuations that appeared to be shear Alfv\'en waves.

\citep{Heyvaerts:1983} and \citep{Similon:1989} showed that when shear Alfv\'en waves propagate along chaotic field lines they are quickly damped.  \color{black}  \citep*{Huang-Bhattacharjee:2022} defined reconnection, even in three-coordinate problem, as the transfer of energy from the reconnecting field to the plasma, not the breaking of field line connections.  Their definition, which is common in two-coordinate models, led them to the odd conclusion that the chaos of the field lines was irrelevant to reconnection---only the formation of localized currents, which are caused by  Alfv\'en wave damping, are relevant.

The rapid spreading of impurities across the entire plasma is a commonly observed feature of tokamak disruptions.  The impurities greatly reduce the damage produced by disruptions by radiating the plasma thermal energy relatively uniformly over the walls rather than having it concentrated in spots where turnstiles formed within the plasma strike the walls as is often the case with runaway electrons.  The reason for the impurity spreading did not have a firm theoretical basis, but the Bohm-like transport that is produced when there is an electron pressure gradient along chaotic field lines, \citep{Boozer:2024:Electric} and Section \ref{Sec:chaos} provide an explanation.  The simulations of \citep{Nardon:2023} saw electric fields consistent in magnitude, but the equation in their code which gave the electric field was not simply related to the two-fluid equations that \citep{Boozer:2024:Electric} used to obtain the Bohm-like transport.

\citep{Breizman:2019} reviewed the problem of the conversion of the plasma current from being carried by near-thermal to being carried by relativistic electrons, which is often viewed as the most dangerous feature of disruptions.   Even though the total energy in the relativistic electrons is only about ten percent of the pre-disruption plasma thermal energy, impurity radiation spreads the thermal energy losses almost uniformly over the wall.  Relativistic electron losses onto the walls can be extremely concentrated in both space and time, which means they are far more destructive.  As pointed out by \citep*{Boozer-Punjabi:2016} the spatial and temporal concentration can be explained by the breaking of the outermost magnetic surface that is confining the runaway electrons in an internal bounded chaotic region.   As the outermost surface breaks it forms a cantorus.  The slower the turnstiles form compared to the timescale for the relativistic electrons to cover the chaotic region the more tightly collimated are the turnstiles on the cantorus.  When the turnstiles form rapidly compared to the timescale for the relativistic electrons to cover the chaotic region, the turnstiles become arbitrary large, which spreads the relativistic electrons so broadly on the walls that they do no damage.  The highly damaging case is often observed and brings fear, but \citep{Reux:2021} and \citep{Paz-Soldan:2021} have observed on the JET and the DIII-D tokamaks the benign spreading of the relativistic electrons when the plasma becomes unstable to a rapidly growing magnetic perturbation.

\citep*{Boozer:2025:Constraints} has pointed out that the frequency with which disruptions occur in tokamaks may be in large part be due to the far larger poloidal flux produced by the plasma current than the change in the poloidal flux that occurs if the profile of the density of the net plasma current $j_{||}$ is varied over the full range of profiles stable to tearing modes.  This flux ratio is approximately a factor of ten.  Equation (\ref{dpsip/dt}), $\partial\psi_p/\partial t=V_\ell$, implies the plasma will disrupt within a shorter time than the time required to shutdown by removing the plasma-produced poloidal flux or the natural length of a flattop pulse unless a spatially constant loop voltage remains consistent with tearing mode stability.  As shown in Equation \ref{Ad-diff}, the topology of the magnetic field is held fixed when $\vec{\nabla}V_{\ell}=0$.  An implication is that poloidal flux can be removed or maintained without producing a disruption when
\begin{equation}
    j_{||} = \frac{V_{\ell}}{2\pi R \eta} + j_{bs}+ j_{cd}
\end{equation}
remains in a non-disruptive profile with $\vec{\nabla}V_{\ell}=0$.  $R$ is the major radius, $\eta$ is the resistivity, which approximately scales as $1/T_e^{3/2}$ but also depends on the charge state of impurities, $Z_{eff}$. \color{black}  The bootstrap current density is $j_{bs}$, and the density of the externally driven current is $j_{cd}$   

The spatial constancy of $V_\ell$ is approximately a constraint on the electron temperature profile to avoid disruptions produced by tearing modes.  The electron temperature profile is determined by microturbulent transport together with the heating and the cooling.  In the shutdown of a fusion plasma, the heating source changes radically as the fusion power turns off.  

The profile of $j_{||}$ can in principle be controlled by plasma heating and current drive.  During the flattop of a fusion pulse, the required power to produce a large change in the $j_{||}$ profile is comparable to the power produced by the alpha-particle heating. This is obvious for direct heating.  \citep*{Boozer:1988} showed the power required to drive the full plasma current in a power plant is greater than $j_{cd} E_{ch}$ integrated over the plasma volume where $E_{ch}$ is the Connor-Hastie electric field \citep{Connor:1975}.  The implication is the power required to drive the total plasma current is comparable to the power produced by alpha-particle heating.  Careful control of the injected power is required to properly control the profile of $j_{||}$, but the available diagnostics may not provide adequate information for such control even if it were in principle possible.  This disruption issue can be circumvented by the use of stellarator magnetic field to produce the rotational transform by external coil currents rather than by the plasma current.


\subsection{Stellarator physics}

Stellarators have a unique external control over the plasma through the coils, which \citep{Boozer:2004} showed could efficiently produce approximately fifty independent magnetic field distributions, an order of magnitude more than can be produced in axisymmetry.  By their nature, stellarators provide steady-state plasma confinement, low recirculating power, as well as robustness against disruption and the loss of positional control within the plasma chamber.

  Adequate stellarator confinement requires optimization of the external field.  In 1980 stellarators were known to have many attractive properties for fusion power plant, but it was thought that this was precluded by the extremely large neoclassical transport caused by their toroidal asymmetry.  \citep{Boozer:1980} gave a new formulation of the drift equations and a new coordinate system that allowed a relatively fast optimization to reduce the neoclassical transport.  \citep*{Nurenberg:1988} used this formalism to show that such magnetic configurations exist. The result of modern optimization for a fusion power plant is illustrated by the Infinity Two fusion pilot plant design \citep{Hegna:2025}.
  

The fifty dimensional space of magnetic fields that can be used for stellarator design is too large for a complete exploration to ever be made, and new stellarator designs are continually being published.  

\citep{Boozer:2024:Stellarators} gave a conceptual design for stellarators with enhanced tritium confinement and edge radiation control.  This design used special features of the stellarator to greatly enhance the tritium burn-up fraction, a non-resonant divertor to allow impurity radiation at the edge to spread the thermal power exhaust broadly over the wall, and  flattop density and temperature profiles to enhance the fusion power production and reduce the impurity confinement.  The low tritium burn-up fraction is a major issue in obtaining tritium self-sufficiency in toroidal fusion power plants.  It should be noted that if a fission source of tritium could be identified that allowed a large fusion to fission power ratio and with a cost of only millions of dollars per kilogram, fusion power plants could be greatly cheapened and simplified by reducing or eliminating their tritium-breeding blankets.  A large fusion/fission power ratio would allow the fission tritium producers to be located at secure sites, such as nuclear weapons facilities.  Poor confinement in the central part of the stellarator would flatten all profiles, including those of impurities, and as discussed in Section \ref{ITG-chaos} the magnetic field chaos produced at non-zero plasma pressure by what is called electrostatic microturbulence could expel impurities.


\section{Discussion \label{Sec:Discussion}}

The incentive for a physicist to learn an area of mathematics is insights on and methods for solving problems of practical importance.   Examples show that the lack of familiarity with the mathematical concepts of magnetic field line chaos, cantori and turnstiles block developments in toroidal plasma physics.  This was the motivation for this review, which is an expanded version of an invited talk at the 29th Workshop on MHD Stability Control in July 2025. 

The importance of field line chaos to the breaking of magnetic field line connections may seem obvious, but the concept receives little notice at international conferences on reconnection.  The work of \citep{Borgogno:2011} on magnetic reconnection is an exception. They considered effects of cantori and turnstiles without using the names—concepts that could have made the generality of their results more apparent.  The ease and rapidity of magnetic reconnection when the magnetic field is chaotic implies what are called electrostatic instabilities can produce non-ambipolar transport through an  electron viscous force.  

The magnetic energy that is released by the breaking of connections in chaotic magnetic fields requires an understanding of the damping of Alfv\'en waves in chaotic magnetic fields, which are the immediate recipients of that energy.  Phase-mixing arguments correctly predict this damping is rapid, but the fraction of the power that is dissipated by resistivity, which is primarily a transfer to the electrons, versus that dissipated by viscosity, which is primarily a transfer to the ions, is not understood.

The primary definition of chaos in the Merriam-Webster dictionary is ``a state of utter confusion,''  but a definition using technical terms is also given: ``the inherent unpredictability in the behavior of a complex natural system.''  Both make the formation of highly collimated flux tubes implausible when the magnetic field is chaotic.  Nevertheless, this collimation is the essential element in the explanation of the extreme localization in space and time of runaway electrons strikes on the chamber walls.  This localization presents the greatest danger to tokamak devices following a disruption.  The successful avoidance of this destructive collimation by \citep{Reux:2021} and \citep{Paz-Soldan:2021} using a rapidly-developing instability is counterintuitive without an understanding of cantori and turnstiles.  

Non-resonant divertors offer the potential for far more control of the plasma edge than do divertors based on a separatrix.  Control is needed for the out-coming particles to follow well-collimated flux tubes into the highly localized ducts  of pumps while defining a sufficiently thick region of poor confinement for high-Z impurities to spread the out-coming power over the walls and and to block in-coming neutrals from charge exchanging within the high energy plasma.  Charge exchange would produce high-energy neutrals that erode the walls, \cite{Boozer:2024:Stellarators}.  Compared to their potential for solving a major issue in the development of fusion power plants, little research is being done on non-resonant divertors, which requires practical methods for controlling chaos, cantori, and turnstiles.

When there is a pressure gradient along magnetic field lines, $\vec{B}\cdot\vec{\nabla}p\neq0$, in addition to the pressure gradient across the field, $\vec{B}\times\vec{\nabla}p$, quasi-neutrality becomes far more subtle than ambipolarity.  Ambipolarity is enforced by a constant electric potential on each magnetic surface.  Although generally ignored in simulations, this subtlety always arises in divertors and during disruptions.   Subtleties exist even without magnetic field line chaos, but with chaos the $\vec{E}\times \vec{B}$ drifts required by quasi-neutrality give transport at a Bohm-like rate.  Practical methods for including these effects in simulations require development, and the realism of effects seen in simulations, which have far less spatial and temporal resolution than is in principle required, remains to be determined.

The literature on the applications of chaos, cantori, and turnstiles in toroidal plasma physics is concentrated among only a few authors, which implies much could be gained from a broader understanding of these concepts.

\color{black}

\section*{Acknowledgements}

This work was supported in part by the U.S. Department of Energy, Office of Science under Award No. DE-AC02-09CH11466).


\appendix

\section{Separation of neighboring lines \label{Sec:separation}}

The behavior of the field lines that are separated by a distance $\rho$ from an arbitrary line $\vec{x}_0(\ell)$ has a general form, \citep*{Boozer:2012}, as $\rho\rightarrow0$.  The trajectories of these infinitesimally separated lines are given in Courant and Snyder intrinsic coordinates by a simple Hamiltonian.  This coordinate system, which is defined in Appendix B of \citep*{Courant:1958}, is $\vec{x}(\rho,\alpha,\ell)=\rho\cos\alpha \hat{\kappa}_0 + \rho \sin\alpha \hat{\tau}_0 + \vec{x}_0(\ell)$, where the  $\hat{b}_0\equiv\vec{B}_0(\ell)/B_0$ along the line, $d\hat{b}_0/d\ell= \kappa_0\hat{\kappa}_0$ is the curvature of the line, and $d\hat{\kappa}_0/d\ell = -(\kappa_0 \hat{b}_0+\tau_0\hat{\tau}_0)$ gives the torsion $\tau_0$.  The torsion measures the extent to which the line fails to lie in a plane.  Letting $\tilde{\psi}\equiv \pi B_0(\ell)\rho^2$, the neighboring trajectories are given by the Hamiltonian 
\begin{eqnarray}
&&\tilde{H}(\tilde{\psi},\alpha,\ell) = \tilde{\psi} h(\alpha,\ell); \\
&& h(\alpha,\ell) =  k_\omega(\ell)+ k_q(\ell)\cos(2\alpha - \varphi_q(\ell)); \\
&& \frac{d\alpha}{d\ell} = \frac{\partial \tilde{H}}{\partial\tilde{\psi} }= h; \\
&& \frac{d\tilde{\psi} }{d\ell} = - \frac{\partial \tilde{H}}{\partial\alpha} = - \tilde{\psi} \frac{\partial h}{\partial \alpha}.  \label{psi-sep}
\end{eqnarray} 
The twist of a field line is given by 
\begin{eqnarray}
k_\omega &\equiv&  \frac{1}{2} K_0(\ell) + \tau_0(\ell), \hspace{0.2in}\mbox{where   } \\
K_0 &\equiv& \frac{\mu_0 j_{||}}{B_0}.
\end{eqnarray}  
The amplitude of the second derivative of $\vec{B}$ with respect to $\rho$ along the line $\vec{x}_0$ gives the quadrapolar term $k_q(\ell)\cos(2\alpha - \varphi_q(\ell))$.


\subsection{Maximum field-line separation \label{Sec:sep} }

Neighboring magnetic field lines to the line $\vec{x}_0(\ell)$ are defined by their position at $\ell=0$, which is given by $\alpha_0$ and $\tilde{\psi}_0$.  There is a maximum separation $\tilde{\psi}$, which is $\tilde{\psi}_0 \exp(\sigma_{max}(\ell))$, and there is a minimum separation  $\tilde{\psi}_0 \exp(-\sigma_{max}(\ell))$ as a function of $\alpha_0$.  To obtain $\sigma_{max}(\ell)$, Equation (\ref{psi-sep}) can be written as $d\tilde{\psi}/d\ell = k_e \tilde{\psi}$, where $k_e\equiv 2k_q(\ell)\sin(2\alpha - \varphi_q(\ell))$ and $\alpha(\ell)$ is given by $d\alpha/d\ell=h(\alpha,\ell)$.  Given a starting point, the function $k_e(\ell)$ can be determined with no knowledge of the dependence of $\tilde{\psi}$ on $\ell$.  The exponential separation of the line is a definite function of $\alpha_0$ and $\ell$ with $\sigma(\alpha_0,\ell) = \int_0^\ell k_e(\ell) d\ell$.  As is clear from the periodic dependence of $k_e$ on  $\alpha$, the function $\sigma(\alpha_0,\ell)$ will have a maximum value as a function of $\alpha_0$ of $\sigma_{max}(\ell)$ and a minimum value equal to minus $\sigma_{max}(\ell)$.  The divergence-free nature of the magnetic field implies there must be equal fluxes approaching the line $\vec{x}_0(\ell)$ as diverging from it.

A non-zero Lyapunov exponent, 
\begin{equation}\lambda_L = \lim_{\ell\rightarrow\infty} \frac{\sigma_{max}(\ell)}{\ell},
\end{equation} 
is sometimes used to define chaos, but $\lambda_L$ can be zero with  $\sigma_{max}(\ell)$ reaching arbitrarily large values.  For reconnection the maximum $\sigma_{max}(\ell)$ is what is relevant not $\lambda_L$, which may be zero.




\subsection{Dependence on only two coordinates \label{Sec:two-coord} }

When the Hamiltonian $\tilde{H}$ depends on only two coordinates, it is not possible for a magnetic field line to come arbitrarily close to every point in a bounded volume, as in the red area of Figure \ref{fig:Stdmap}.

This is illustrated by letting $\tilde{H}$ be a function of only $\alpha_q = \alpha - \ell/R_q$ and $\tilde{\psi}$.  A canonical transformation gives a Hamiltonian
\begin{eqnarray}
\tilde{\mathcal{H}}(\tilde{\psi},\alpha_q) &=& \Big(K_\omega + k_q \cos(2\alpha_q) \Big) \tilde{\psi}, \\
K_\omega &\equiv& k_\omega - \frac{1}{R_q}
\end{eqnarray}
with $K_\omega$ and $k_q$ constants.  The correctness of the evolution equations $d\alpha_q/d\ell = \partial\tilde{\mathcal{H}}/\partial\tilde{\psi}$ and $d\tilde{\psi}/d\ell = -\partial\tilde{\mathcal{H}}/\partial\alpha_q$ is easily checked.  The generating function $G=(\alpha-\ell/R_q)\tilde{\psi}$ transforms the canonical coordinate $\alpha$ to $\alpha_q = \partial G/\partial\tilde{\psi}$, transforms the Hamiltonian to $\tilde{\mathcal{H}} = \tilde{H}+ \partial G/\partial\ell$ or $\tilde{\mathcal{H}} =  \tilde{H} - \tilde{\psi}/R_q$, and leaves the canonical momentum $\tilde{\psi}$ unchanged.

The change in $\tilde{\mathcal{H}}$ along a magnetic field line is $d\tilde{\mathcal{H}}/d\ell=0$.   Since $\tilde{\mathcal{H}}$ is constant along a field line, it would have to be constant throughout a volume filled by a single magnetic field line, as in the red area of Figure \ref{fig:Stdmap}, which gives the contradiction that $d\alpha_q/d\ell$ and $d\tilde{\psi}/d\ell$ would need to be zero as well.  \\


\subsection{Required current density for a given $\sigma_{max}(\ell)$ \label{Sec:j-max} }

Since $k_e$ is linearly dependent on the magnetic field, the required current density to produce that field is only linearly dependent on the current density.  A curl-free field can have an arbitrarily large $\sigma_{max}(\ell)$, but an ideal evolution can cause an arbitrarily large increase in $\sigma_{max}(\ell)$ with a required current density that scales linearly with $\sigma_{max}(\ell)$.  




\bibliographystyle{jpp}

\bibliography{jpp-instructions}

\end{document}